\documentclass[aps,pre,twocolumn,amsmath,amssymb,floatfix,showpacs]{revtex4}

\usepackage[dvipsnames,rgb,dvips]{xcolor}
\usepackage{overpic}

\usepackage{graphicx,mathrsfs,psfrag}
\usepackage{dcolumn}

\newcommand{\Reys}{\ensuremath{\textrm{Re}_s}}
\newcommand{\St}{\ensuremath{\textrm{St}}}

\renewcommand{\cite}{\citep}
\renewcommand{\tilde}{\widetilde}
\renewcommand{\cite}{\citep}
\newcommand{\ve}[1]{\ensuremath{\mbox{\boldmath$#1$}}}

\newcommand\nn{\nonumber}

\renewcommand{\vec}{\mathbf}
\newcommand{\tens}[1]{\boldsymbol{\mathsf{#1}}}

\begin{document}
\title{The role of inertia for the rotation of a nearly spherical particle in a general linear flow}
\author{F. Candelier$^{1}$, J. Einarsson$^{2}$, F. Lundell$^{3}$, B. Mehlig$^{2}$ and J.-R. Angilella$^{4}$ }
\affiliation{\mbox{}$^{1)}$University of Aix-Marseille, CNRS, IUSTI UMR 7343, 
13 013 Marseille, Cedex 13, France}
\affiliation{\mbox{}$^{2)}$Department of Physics, Gothenburg University, SE-41296 Gothenburg, Sweden}
\affiliation{\mbox{}$^{3)}$KTH Royal Institute of Technology, SE-100 44 Stockholm, Sweden }
\affiliation{\mbox{}$^{4)}$Department of Mathematics and Mechanics, LUSAC-ESIX, University of Caen, France}

\begin{abstract}
We analyse the angular dynamics of a neutrally buoyant  nearly spherical 
particle immersed in a steady general linear flow.
The hydrodynamic torque acting on the particle is obtained by means of a reciprocal theorem,
regular perturbation theory exploiting the small eccentricity  of the nearly spherical particle, and assuming
that inertial effects are small, but finite. 
\end{abstract}
\maketitle

\section{Introduction}
In this article we derive an effective equation of motion for the orientational dynamics of a neutrally buoyant, 
nearly spherical axisymmetric particle suspended in a time-independent linear flow. 
{Our result is valid to leading order in the shear Reynolds number $\Reys$ and the particle eccentricity $\epsilon$.
Terms of order $\Reys^2$, $\epsilon^3$, and $\epsilon^2 \Reys$ are neglected.   

Our motivation was two-fold. Firstly, we have recently computed the stability of log-rolling and tumbling orbits of a neutrally buoyant spheroid in a
 simple shear at weak fluid and particle inertia \cite{einarsson2015a,einarsson2015b}. 
The calculations leading  to these results are quite involved.
We therefore decided to check our calculations by an alternative method, summarised below. 
We refer to Ref.~\cite{einarsson2015b} for a summary of the background of the problem, and for a discussion
of the implications of the results.
Secondly, the results 
described in the present article are valid for general linear flows while those given in Refs.~\cite{einarsson2015a,einarsson2015b}
pertain to the particular (and important) case of a simple shear flow. We believe that it is of interest to obtain results for general linear 
flows because this can be a first step towards describing the effect of a  time-dependent but slowly varying perturbation of the flow.

\section{Formulation of the problem} We consider a nearly spherical particle corresponding to an ellipsoid of revolution (around the axis $\vec{t}_1$) of low eccentricity as depicted in Fig.~\ref{fig:1}.  
The surface of the particle is parametrised as
\begin{equation}
r(\vartheta) = 1 -\epsilon \sin^2(\vartheta) -  \frac{3}{2} \:\epsilon^2 \:\sin^2(\vartheta) \cos^2(\vartheta) + O(\epsilon^3)\:.
\label{eq:shape}
\end{equation}
Here $\vartheta$ denotes the polar angle made by any vector $\vec{r}$ with the orientation of the axis of revolution $\vec{t}_1$.
The eccentricity $\epsilon$ is a small parameter. Lengths are normalized by the semi-axis length $a$ along the direction $\vec{t}_1$
(Fig. \ref{fig:1}).

Suppose that the particle is immersed in a steady general linear flow. 
In this case the angular dynamics of the particle is governed by  three dimensionless parameters: 
the shear Reynolds number $\Reys = s a^2 \rho_{\rm f}/\mu$ which measures the effect of fluid inertia, 
the Stokes number $\mbox{St} = (\rho_{\rm p}/\rho_{\rm f}) \Reys$ measuring the importance of particle inertia,
and the particle eccentricity $\epsilon$. Here $\rho_{\rm f}$ and $\rho_{\rm p}$ denote, respectively, the densities 
of the fluid and of the particle, $s$ denotes the shear rate of the linear flow, and $\mu$ the dynamic viscosity of the fluid. 
\begin{figure}[t]
\includegraphics{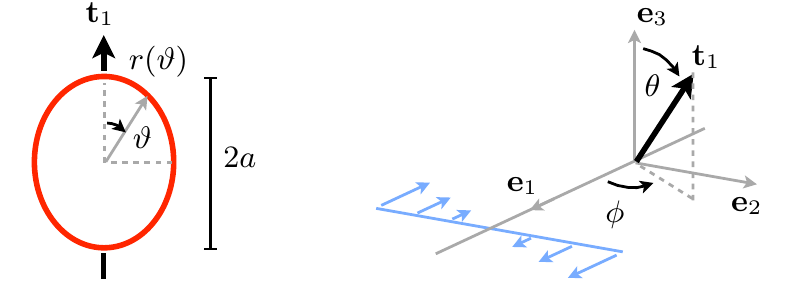}
\caption{\label{fig:1} {\em Color online.} Left: the surface of a nearly spherical particle is constructed as the surface of revolution 
of the curve Eq.~(\ref{eq:1}). Right: simple shear. The flow-shear plane is spanned by ${\bf e}_1$ and ${\bf e}_2$, and the vorticity
points in the negative ${\bf e}_3$-direction. The orientation of the vector ${\bf t}_1$ pointing along the symmetry axis of the particle is expressed in spherical coordinates $(\theta,\phi)$.}
\end{figure}

Using the inverse shear rate as the time scale, the semi-axis length  $a$ as the length scale and $\mu s$ 
as the pressure scale,  the equations governing the angular dynamics of a neutrally buoyant particle read in the laboratory frame of reference
\begin{equation}
\frac{ \mbox{d} \vec{t}_1}{\mbox{d} t}  = \boldsymbol{\omega} \times  \vec{t}_1 \:,
\label{eq:constraint}
\end{equation}
and
\begin{equation}
\mbox{St} \frac{ \mbox{d}  }{\mbox{d} t}\Big(\tens{J} \cdot  \boldsymbol{\omega}\Big) = \boldsymbol{\tau}_h\,.
\label{eq:1}
\end{equation}
In Eq. (\ref{eq:1}), $\tens{J}$ denotes the moment-of-inertia tensor of the particle, $ \boldsymbol{\omega}$ its angular velocity, 
and $\boldsymbol{\tau}_h$ is the hydrodynamic torque acting on the particle. Throughout this paper the centred dot  ($\cdot$)   defines a simply contracted tensor product, uppercase letters are used to denote matrix-tensors and lowercase letters are used for  simple vectors. 

For a neutrally buoyant particle we have $\St =\Reys$. Making this substitution only at the end of the calculation allows us to separate the effects of particle and fluid inertia.

The hydrodynamic torque is given by 
$$
\boldsymbol{\tau}_h = \int_\mathscr{S} \vec{r} \times \Big( \boldsymbol{\Sigma} \cdot \vec{n} \Big) \mbox{d} S\,.
$$
The integral is over the particle surface $\mathscr{S}$, and $\vec{n}$ is the outward surface normal. 
For an incompressible Newtonian fluid the stress tensor reads 
$\boldsymbol{\Sigma} = -p \:\tens{I} + 2\:\tens{S}$ \,,
where $p$ is the pressure, $\tens{I}$ the identity tensor and $\tens{S}$ the symmetric part of the fluid-velocity gradient tensor. 
To determine the hydrodynamic torque acting on the spheroid, we  use a reciprocal theorem 
\cite{lorentz,happel1983,kim1991,subramanian2005,subramanian2006}. 

\section{Method} 
We consider  a general steady linear ambient flow. In the laboratory frame, it takes  the form 
$$
\vec{u}^\infty = \tens{A}^\infty \cdot \vec{x}\:,
$$
where 
$\tens{A}^\infty$ a constant tensor. We decompose the gradient matrix $ \tens{A}^\infty$
into its symmetric ($\tens{S}^\infty$) and antisymmetric ($\tens{O}^\infty$) parts:
$$
\tens{A}^\infty = \tens{O}^\infty +  \tens{S}^\infty\:.
$$
The anti-symmetric tensor $\tens{O}^\infty$ is linked to the vorticity of the unperturbed flow through
$$
 \tens{\tens{O}^\infty}\cdot \vec{x} = \:\boldsymbol{\omega}^\infty \times \vec{x} \quad   \mbox{where} \quad \boldsymbol{\omega}^\infty  = \frac{1}{2}  \:\boldsymbol{\nabla}\times \vec{u}^\infty \,.
$$
The vectors $\vec{x}_p$ and $\vec{u}_p$  
denote the the position and the velocity of the centre-of-mass of  the particle. 
As the spheroid is assumed to be neutrally buoyant, we consider it to be advected along streamlines with a vanishing slip velocity 
$$
\vec{u}_p = \tens{A}^\infty \cdot \vec{x}_p\:.
$$
We consider the equations governing the fluid motion in a translating frame of reference centred on the particle. This frame being non-inertial, the pseudo-force $-\Reys {\rm d}{\ve u}_p/{\rm d}t$ appears in the force balance acting on the particle. 
However, it does not appear in the equations of the perturbation flow below.
In the moving frame of reference the Navier-Stokes equations for the fluid velocity $\vec{u}$ read:
$$
\boldsymbol{\nabla} \cdot  \vec{u} = 0\:,
$$
\begin{equation}
\mbox{Re}_s \:\left(\frac{\partial \vec{u}} {\partial t} + (\vec{u} \cdot\boldsymbol{\nabla} ) \vec{u} \right) = \boldsymbol{\nabla} \cdot \boldsymbol{\Sigma} - \mbox{Re}_s   \frac{\mbox{d} \vec{u}_p}{\mbox{d}t}\,,
\label{eq:NS1}
\end{equation}
with  boundary conditions 
$$
\vec{u} = \boldsymbol{\omega} \times \vec{r}\,,
 \quad \vec{r}  \in \mathscr{S} \quad \mbox{and} \quad \vec{u} \to \tens{A}^\infty \cdot \vec{r} \,, \quad r \to \infty\,,
$$
with $\vec{r} = \vec{x}-\vec{x}_p$.

We use the reciprocal theorem to determine the torque acting on the spheroid. 
The reciprocal theorem relates integrals of the velocity and stress fields of the two incompressible Newtonian fluids:
\begin{align}
& \int_\mathscr{S}\!\! \Big((\tilde{\boldsymbol{\Sigma}} \cdot \vec{n} ) \cdot  \vec{u}  - (\boldsymbol{\Sigma} \cdot \vec{n}) \cdot  \tilde{\vec{u}} \Big) \mbox{d}S \label{eq:rt}\\
&\qquad
  \!=\!\!\int_\mathscr{V} \!\!\Big(\tilde{\vec{u}} \cdot (\boldsymbol{ \nabla} \cdot \boldsymbol{\Sigma})-\vec{u} \cdot (\boldsymbol{ \nabla} \cdot  \tilde{\boldsymbol{\Sigma}}) \Big) \mbox{d} V \,,\nn
\end{align}
where the integral on the  r.h.s is over the entire volume $\mathscr{V}$ outside the particle. In Eq. (\ref{eq:rt})  the set 
($\vec{u},\,\boldsymbol{\Sigma}$) represents the solutions of the problem of interest and 
($\tilde{\vec{u}},\:\tilde{\boldsymbol{\Sigma}}$)  
are the solutions of a suitable auxiliary problem describing 
the creeping flow produced by
a particle moving with angular velocity $\tilde{\boldsymbol{\omega}}$ in an otherwise quiescent fluid. 

We also introduce the decompositions
 $$
 \vec{u} = \tens{A}^\infty  \cdot \vec{r} + \vec{u}' \quad \mbox{and} \quad \boldsymbol{\Sigma}  =  \boldsymbol{\Sigma}^\infty +  \boldsymbol{\Sigma}'\:,
 $$
 where $\vec{u}'$ and $\boldsymbol{\Sigma}'$ correspond to the perturbation flow and the perturbation stress tensor induced by the spheroid. 
 Using these decompositions it follows from the boundary conditions for  $\vec{u}'$ both on the particle surface and at  infinity that:
 \begin{align}
\boldsymbol{\omega}_s  \cdot \tilde{\boldsymbol{\tau}}_h& - \int_\mathcal{S} \left(\tilde{\boldsymbol{\Sigma}} \cdot \vec{n}\right) \cdot \left(\tens{S}^\infty \cdot \vec{r} \right) \:\mbox{d}{S} \label{eq:rt3}\\
&= \tilde{\boldsymbol{\omega}} \cdot \boldsymbol{\tau}_h + \mbox{Re}_s \int_\mathcal{V} \vec{f}(\vec{u}')  \cdot \tilde{\vec{u}} \:\mbox{d} {V}\nn\,.
\end{align}
In this equation we have introduced the slip angular velocity 
$$
\boldsymbol{\omega}_s = \boldsymbol{\omega} - \boldsymbol{\omega}^\infty\:, 
$$
and  the function $\vec{f}(\vec{u}')$ in Eq.~(\ref{eq:rt3}) is given by
$$
\vec{f}(\vec{u}') = \frac{\partial \vec{u}'}{\partial t} + 
\tens{A}^\infty\cdot \vec{u}'+ 
\Big((\tens{A}^\infty\cdot \vec{r}) \cdot \boldsymbol{\nabla} \Big) \vec{u}' + 
\vec{u}'\cdot \boldsymbol{\nabla} \vec{u'}\:.
$$
To derive Eq. (\ref{eq:rt3}) we have also used the fact that the torque $\boldsymbol{\tau}_h^\infty$ due to the unperturbed stress vanishes. To see this recall that in the laboratory frame of reference, the unperturbed flow field satisfies the Navier-Stokes equations $\tens{A}^2 \cdot \vec{x} = \boldsymbol{\nabla} \cdot \boldsymbol{\Sigma}^\infty$. Using Stokes integration theorem it follows that 
the torque due to the unperturbed stress tensor is given by 
\begin{equation}
\label{eq:thinf}
\boldsymbol{\tau}_h^\infty = \int_{\mathscr{V}_p} \vec{r} \times \Big({(\tens{A}^\infty)}^2 \cdot \vec{x}_p\Big) \: \mbox{d}V  +  \int_{\mathscr{V}_p} \vec{r} \times \Big({(\tens{A}^\infty)}^2 \cdot \vec{r}\Big) \: \mbox{d}V \,,
\end{equation}
where the integral is over the volume {$\mathscr{V}_p$} inside the particle. 
The symmetrical shape of the particle implies that the first integral on the r.h.s of Eq. (\ref{eq:thinf}) is zero.  Incompressibility 
and the vorticity equation for a steady flow
\begin{equation}
\mbox{Tr}(\tens{S}^\infty) = \vec{0} \quad \mbox{and}  \quad \tens{S}^\infty \cdot \boldsymbol{\omega}^\infty  = 0
\label{eq:identities}
\end{equation}
imply that the second integral also vanishes. 

Eq.~(\ref{eq:rt3}) allows to compute the hydrodynamic torque $\boldsymbol{\tau}_h$ 
provided that we can determine the contribution of the integral over the entire volume $\mathscr{V}$ outside the particle.
To leading order \cite{einarsson2015b,subramanian2006} one can replace the actual perturbation flow  $\vec{u}'$ in the integral
by its limit of vanishingly small $\mbox{Re}_s$. In other words, $\vec{u}'$ can be 
replaced by the solution of
\begin{align}
0&=\boldsymbol{\nabla} \cdot \vec{u}' \:, \quad 0 =-\boldsymbol{\nabla} p' + \boldsymbol{\triangle } \vec{u}'  
\label{eq:Stokes1}
\end{align}
with boundary conditions
\begin{align}
\vec{u}' &= (\boldsymbol{\omega} - \boldsymbol{\omega}^\infty) \times  \vec{r} - \tens{S}^\infty\cdot  \vec{r}\quad\mbox{for}\quad \vec{r}  \in \mathscr{S}\,,\nn\\
\vec{u}' &\to \vec{0} \,\quad\mbox{as}\quad  r \to \infty\,.
\label{eq:Stokes2}
\end{align}
To determine $\vec{u}'$  we follow Refs.~\cite{Brenner1964,Hinch1991} and use 
a perturbation method that assumes  that the particle eccentricity $\epsilon$ is small. 
Our approach is based on the general Lamb solution \cite{lamb1945} of the Stokes problem (\ref{eq:Stokes1},\ref{eq:Stokes2})  derived in terms of spherical harmonics. Details are given in the appendix. 

\section{Results for general linear flows}
{Eq.~(\ref{eq:rt3}) makes it possible to} derive the torque acting on a neutrally buoyant spheroid immersed in a general linear flow. 
For the auxiliary torque -- the first term on the l.h.s. of Eq. (\ref{eq:rt3}) --
we find:
\begin{align}
\tilde{\boldsymbol{\tau}}_h& = -8\pi \Big(1-\frac{9}{5} \epsilon +\frac{459}{350} \epsilon^2\Big) \tilde{\boldsymbol{\omega}}\nn\\
&+8\pi \Big(\frac{3}{5} \epsilon - \frac{177}{350} \epsilon^2\Big) \tilde{\Omega}^1 \:\vec{t}_1 + O(\epsilon^3)\,.\nn
\end{align}
The second term on the l.h.s. of Eq. (\ref{eq:rt3}) evaluates to:
\begin{align}
&\int_\mathcal{S} \left(\tilde{\boldsymbol{\Sigma}} \cdot \vec{n}\right) \cdot \left(\tens{S}^\infty \cdot \vec{r} \right) \:\mbox{d}\mathcal{S}\nn\\
&\qquad= 8\pi \Big(\epsilon - \frac{13}{10} \epsilon^2\Big)\Big( (\tens{S}^\infty \cdot \vec{t}_1)\times \vec{t}_1 \Big)\cdot \tilde{\boldsymbol{\omega}} + O(\epsilon^3)\,. \nn
\end{align}
In order to determine the second term on the r.h.s of Eq. (\ref{eq:rt3}), which is, as already said, the most difficult part of this investigation, we first make use of the fact that the slip angular velocity is expected 
to be of order $O(\epsilon)$. Indeed, Jeffery's theory 
\cite{jeffery1922} predicts        
$$
\lim_{\mbox{Re}_s\to 0} \boldsymbol{\omega}_s = \Lambda \,\vec{t}_1\: \times (\tens{S}^\infty \cdot \vec{t}_1)\:.
$$
The factor $\Lambda$ is  a shape factor determined by the particle aspect ratio $\lambda$ (the ratio between the length 
of the particle along its symmetry axis and its length transverse the symmetry axis)  through 
$$
\Lambda =  \frac{\lambda^2-1}{\lambda^2+1} \sim \epsilon + \frac{\epsilon^2}{2} + O(\epsilon^3) \,.
$$
Using the fact that the slip angular velocity
is of order $O(\epsilon)$ and making use of the identities Eq.~(\ref{eq:identities}) we find 
\begin{align}
& \mbox{Re}_s \int_\mathcal{V} \vec{f}(\vec{u}')  \cdot \tilde{\vec{u}} \:\mbox{d} \mathcal{V} \nn\\
 &\qquad=
  -8\pi \mbox{Re}_s \Big(  -\frac{1}{3} \frac{\mbox{d}  \boldsymbol{\omega}_s}{\mbox{d}t}  + \frac{1}{8}\: \tens{S}^\infty \cdot \boldsymbol{\omega}_s \Big) \cdot \boldsymbol{\tilde{\omega}} \nn\\
 &\qquad\quad -8 \pi \epsilon \mbox{Re}_s \Big(- \frac{32}{105} \big((\tens{S}^\infty\cdot \vec{t}_1)\cdot\vec{t}_1\big) \boldsymbol{\omega}^\infty\nn \\
 &  \qquad\quad +  (\tens{M}\cdot \vec{t}_1)\times \vec{t}_1 \Big) \cdot \boldsymbol{\tilde{\omega}}  + O(\epsilon^2 \mbox{Re}_s)\:, 
\nn
\end{align}
 where 
\begin{equation}
\tens{M}  =    -\frac{3}{5} \tens{S}^\infty \cdot {\tens{O}^\infty}  +\frac{3}{35} {\tens{O}^\infty} \cdot \tens{S}^\infty - \frac{123}{280}
 \tens{S}^\infty \cdot  \tens{S}^\infty  \,.
\end{equation}
So far the hydrodynamic torque  acting on the spheroid in a general linear flow is given in a form involving the scalar product with 
the angular velocity $\boldsymbol{\tilde{\omega}}$ {of} the auxiliary problem. Since these results are valid 
{for arbitrary} $\boldsymbol{\tilde{\omega}}$ we conclude 
\begin{widetext}
\begin{equation}
\begin{split}
\boldsymbol{\tau}_h  = &  -8\pi \Big(1-\frac{9}{5} \epsilon +\frac{459}{350} \epsilon^2\Big)\boldsymbol{\omega}_s
+8\pi \Big(\frac{3}{5} \epsilon - \frac{177}{350} \epsilon^2\Big) (\boldsymbol{\omega}_s \cdot \vec{t}_1) \:\vec{t}_1 
-8\pi \Big(\epsilon - \frac{13}{10} \epsilon^2\Big)\left( (\tens{S}^\infty \cdot \vec{t}_1)\times \vec{t}_1 \right) \\
 &  +8\pi \Reys\Big( -\frac{1}{3} \frac{\mbox{d}  \boldsymbol{\omega}_s}{\mbox{d}t}  + \frac{1}{8}\: \tens{S}^\infty \cdot \boldsymbol{\omega}_s\Big) 
 +8 \pi \epsilon \Reys \Big(- \frac{32}{105} \big((\tens{S}^\infty\cdot \vec{t}_1)\cdot\vec{t}_1\big) \boldsymbol{\omega}^\infty 
 +  (\tens{M}\cdot \vec{t}_1)\times \vec{t}_1 \Big) 
+O(\epsilon^2\Reys)\,.
\label{eq:torque}
 \end{split}
\end{equation}
\end{widetext}
This equation constitutes one of the main results of this paper. Note that the terms in the first row of the r.h.s of Eq. (\ref{eq:torque}) 
{correspond to the creeping-flow limit $(\Reys=0$)}.
The terms in the second row account for fluid-inertia effects. {Since} the slip angular velocity scales as $O(\epsilon)$ 
{we infer} that these effects scale as $O(\epsilon \mbox{Re}_s)$ {for any linear flow.}
 
 In what follows,  an approximate dynamical equation for the angular {dynamics} of the particle is derived. To do so, we need to return to  Eq. (\ref{eq:1}) which governs the orientational dynamics of the spheroid. 
 The moment-of-inertia tensor of the nearly spherical particle reads 
$$
\tens{J} = A^I \vec{t}_1 \otimes \vec{t}_1 + B^I ( \vec{t}_2 \otimes \vec{t}_2+ \vec{t}_3 \otimes \vec{t}_3)
$$
where
\begin{align}
A^I &= \frac{8 \pi }{15}(1-4\epsilon+6 \epsilon^2) + O(\epsilon^3)\:,\nn\\
 B^I &= \frac{8 \pi }{15}(1-3\epsilon+\frac{7}{2}\epsilon^2) + O(\epsilon^3) \:.\nn
\end{align}
Expanding the time derivative in  Eq.~(\ref{eq:1}) yields
$$
\mbox{St} \frac{ \mbox{d}  }{\mbox{d} t}\Big(\tens{J} \cdot  \boldsymbol{\omega}\Big) = \mbox{St} \big(\tens{J} \cdot  \frac{\mbox{d} \boldsymbol{\omega}}{\mbox{d}t} + (A^I-B^I) (\boldsymbol{\omega} \times \vec{t}_1) (\boldsymbol{\omega \cdot \vec{t}_1}) \big)\:.
$$
Writing this equation in terms of the slip angular velocity, and making use of the fact that ${\mbox{d} \boldsymbol{\omega}^\infty}/{\mbox{d}t}= 0$,  we are led to 
\begin{equation}
\begin{split}
\mbox{St} \frac{ \mbox{d}  }{\mbox{d} t}\Big(\tens{J} \cdot  \boldsymbol{\omega}\Big)  & = 8 \pi \mbox{St} \Big( \frac{1}{15} \frac{\mbox{d} \boldsymbol{\omega}_s}{\mbox{d}t}  \Big) \\
 & - 8\pi \epsilon \mbox{St} \Big(  
 \frac{1}{15}  ( \boldsymbol{\omega}^\infty \cdot \vec{t}_1) (  \boldsymbol{\omega}^\infty \times \vec{t}_1)  \Big)\\
 & +O\big(\epsilon^2 \mbox{St}\big)\:. 
 \label{eq:ang_acceleration}
\end{split}
\end{equation}
From Eqs.~(\ref{eq:1}), (\ref{eq:torque}), and (\ref{eq:ang_acceleration}) we obtain
\begin{widetext}
\begin{align}    
\nn
\vec{0}& = \Big(1-\frac{9}{5} \epsilon +\frac{459}{350} \epsilon^2\Big)\boldsymbol{\omega}_s
- \Big(\frac{3}{5} \epsilon - \frac{177}{350} \epsilon^2\Big) (\boldsymbol{\omega}_s\cdot \vec{t}_1) \:\vec{t}_1 
+ \Big(\epsilon - \frac{13}{10} \epsilon^2\Big)\left( (\tens{S}^\infty \cdot \vec{t}_1)\times \vec{t}_1 \right) 
+\mbox{St} \Big( \frac{1}{15} \frac{\mbox{d} \boldsymbol{\omega}_s}{\mbox{d}t}  \Big) \\
 &  + \mbox{Re$_{\rm s}$}\Big( \frac{1}{3} \frac{\mbox{d} \boldsymbol{\omega}_s}{\mbox{d}t}  
 - \frac{1}{8}\: \tens{S}^\infty \cdot \boldsymbol{\omega}_s\Big)
  - \epsilon \mbox{St} \Big[
  \frac{1}{15}  ( \boldsymbol{\omega}^\infty \cdot \vec{t}_1) \Big(  \boldsymbol{\omega}^\infty \times \vec{t}_1 \Big)  \Big]
  + \epsilon \mbox{Re$_{\rm s}$} \Big[ \frac{32}{105} \Big((\tens{S}^\infty\cdot \vec{t}_1)\cdot\vec{t}_1\Big) \boldsymbol{\omega}^\infty - (\tens{M}\cdot \vec{t}_1)\times \vec{t}_1\nn
\Big] \\
& +O\Big(\epsilon^2 \mbox{St},\:\epsilon^2 \mbox{Re$_{\rm s}$}\Big)\,.
\label{eq_main}
\end{align}
\end{widetext}
From this equation we compute the particle angular velocity order by order. To this end we insert the ansatz
\begin{align}
\label{eq:w0}
\boldsymbol{\omega}_s &=  \boldsymbol{\omega}_s^{(0)} \!+\! \epsilon \: \boldsymbol{\omega}_s^{(\epsilon)} \! +\!\mbox{St}  \: \boldsymbol{\omega}_s^{(\mbox{\scriptsize St})}  \!+\!\mbox{Re$_{\rm s}$}  \: \boldsymbol{\omega}_s^{(\mbox{\scriptsize Re$_{\rm s}$})} \!+\! \epsilon^2 \: \boldsymbol{\omega}_s^{(\epsilon^2)} \\
&  + \epsilon \mbox{St}  \: \boldsymbol{\omega}_s^{(\epsilon \mbox{\scriptsize St})}\! + \!\epsilon \mbox{Re$_{\rm s}$}  \: \boldsymbol{\omega}_s^{(\epsilon\mbox{\scriptsize Re$_{\rm s}$})}\! +\! O\Big(\epsilon^3,\epsilon^2 \mbox{St},\:\epsilon^2 \mbox{Re$_{\rm s}$}\Big)\:,\nn
\end{align}
into Eq.~(\ref{eq_main}). To order $O(1)$ this gives
$$
\boldsymbol{\omega}_s^{(0)}= \vec{0}\,.
$$
This means that to leading order the angular velocity of the particle is governed by the vorticity of the unperturbed flow ($\boldsymbol{\omega}^{(0)} = \boldsymbol{\omega}^\infty$). 
To order $O(\epsilon)$ we find:
\begin{equation}
\boldsymbol{\omega}_s^{(\epsilon)}  = \vec{t}_1\times \big(\tens{S}^\infty\cdot \vec{t}_1\big)\:.
\label{eq:w1}
\end{equation}
This term is the leading-order term of Jeffery's angular velocity. 
It combines
with the order-$\epsilon^2$ term
\begin{equation}
\boldsymbol{\omega}_s^{(\epsilon^2)} = \frac{1}{2} \: \vec{t}_1\times \big(\tens{S}^\infty\cdot \vec{t}_1\big)
\label{eq:w2}
\end{equation}
to give Jeffery's angular velocity  $\Lambda \: \vec{t}_1\times \big(\tens{S}^\infty\cdot \vec{t}_1\big)$ to order $O(\epsilon^3)$. To orders  $O(\mbox{St})$ and $O(\mbox{Re$_{\rm s}$})$ we find:
$$
\boldsymbol{\omega}_s^{(\mbox{\scriptsize St})} = \vec{0} \,,\quad
\boldsymbol{\omega}_s^{(\mbox{\scriptsize Re$_{\rm s}$})} = \vec{0}\,.
$$
These results suggest that for any linear flow, and up to first order, neither particle inertia nor fluid inertia modify the orientation of the angular velocity of a sphere as {is} expected from  symmetry arguments. 
Finally, to  orders $O(\epsilon \mbox{St})$ and $O(\epsilon \mbox{Re$_{\rm s}$})$ we find
\begin{equation}
\begin{split}
\label{eq:w3}
\boldsymbol{\omega}_s^{(\epsilon \mbox{\scriptsize St})} 
= & -\frac{1}{15}\big[  (\tens{O}^\infty \cdot \vec{t}_1)\times (\tens{S}^\infty\cdot\vec{t}_1) \\
&+ \vec{t}_1 \times (\tens{S}^\infty\cdot \tens{O}^\infty \cdot \vec{t}_1) \\
& -  (\boldsymbol{\omega}^\infty\cdot\vec{t}_1)\:\: \tens{O}^\infty \cdot \vec{t}_1\big]\:,
\end{split}
\end{equation}
and
\begin{align}    
\label{eq:w4}
&\boldsymbol{\omega}_s^{(\epsilon \mbox{\scriptsize Re$_{\rm s}$})} 
= -\frac{1}{3} (\tens{O}^\infty \cdot \vec{t}_1)\times (\tens{S}^\infty\cdot \vec{t}_1)\\
& \quad\!\!- \frac{1}{3} \vec{t}_1 \times \big(\tens{S}^\infty\cdot \tens{O}^\infty \cdot \vec{t}_1\big) \!+\! \frac{1}{8} \tens{S}^\infty\cdot \big(\vec{t}_1 \times (\tens{S}^\infty\cdot \vec{t}_1) \big)\nn\\
&\quad\!\!- \frac{32}{105} \big((\tens{S}^\infty\cdot \vec{t}_1)\cdot\vec{t}_1\big)  \boldsymbol{\omega}^\infty \nn\\
&\quad\!\!-\Big[\Big( \frac{3}{5} \tens{S}^\infty \cdot{\tens{O}^\infty}  
\!-\!\frac{3}{35} {\tens{O}^\infty} \cdot\tens{S}^\infty  \!+ \!\frac{123}{280}
 \tens{S}^\infty \cdot\tens{S}^\infty \Big) \cdot \vec{t}_1 \Big] \times \vec{t}_1\,.\nn
\end{align} 
These correction terms account for particle-inertia and fluid-inertia {effects}. 
The particle-inertia contribution to the particle angular velocity was calculated in Ref.~\cite{Ein14} for a spheroid immersed in a general linear flow. Expanding Eq. (8) in Ref. \cite{Ein14} to first order in $\epsilon$ results in an expression consistent with Eq. (\ref{eq:w3}).

From Eqs.~(\ref{eq:w3}) and (\ref{eq:w4}) we see that both correction terms are quadratic in the ambient flow-gradient tensors $\tens{A}^\infty$.
We also observe that the orientation vector $\vec{t}_1$ occurs twice in each part composing these corrections. 
{This implies} that the slip angular velocity remains unchanged {upon replacing} $\vec{t}_1$ by $-\vec{t}_1$ 
(particle inversion symmetry).

The above results for the angular velocity of the particle give rise to an effective equation of motion
that allows to examine the role played by inertial effects on the rotation of a 
nearly spherical particle in general linear flows. To this end
we parametrise the orientation of the particle using the polar
angle $\theta$ and the azimuthal angle $\phi$, see Fig.~\ref{fig:1}. In the 
laboratory basis $\vec{e}_i$ the orientation vector $\vec{t}_1$ reads
\begin{equation}
\vec{t}_1 = \sin \theta \cos \phi \,\vec{e}_1 + \sin \theta \sin \phi \, \vec{e}_2 +
 \cos \theta \, \vec{e}_3\,.
\label{eq:te}
\end{equation}
Eqs. (\ref{eq:constraint}) and (\ref{eq:w0}) -- (\ref{eq:te})
result in a non-linear dynamical system of the form
\begin{equation}
\dot{\phi} = f(\theta,\,\phi)\,, \quad \dot{\theta} = g(\theta,\,\phi) \:.
\label{eq:result}
\end{equation}
In the following section we discuss the explicit forms of this equation for three
different linear flows.

\section{Examples} 
{\em Pure shear flow.} 
For a pure shear flow
the matrix $\tens{A}^\infty$  in the laboratory basis  reads
$$ 
\tens{A}^\infty = \left( \begin{array}{ccc} 0 & 1 & 0 \\ 
0 & 0 & 0\\
0& 0& 0\\
 \end{array} \right)\:,
 $$
see Fig.~\ref{fig:1}. Eq.~(\ref{eq:result}) takes the form
\begin{widetext}
\begin{subequations}
\label{eq:eom_small_eps}
\begin{align}    
\dot{\phi} = &\frac{1}{2} ({\epsilon} \cos 2\phi - 1) - \frac{\sin 2\phi}{4} \Big[ \Big(\frac{\epsilon \mbox{St}}{15} + \frac{\epsilon \Reys}{35} \Big) \sin^2 \theta  +  \frac{\epsilon \mbox{St}}{15} - \frac{37 \epsilon \Reys}{105} \Big]\,,\\
\dot{\theta} = & {\epsilon} \frac{\sin 2 \theta  \sin 2 \phi }{4} + \frac{\sin \theta \cos \theta }{4} 
\Big[  \Big(\frac{\epsilon \mbox{St}}{15} - \frac{37 \epsilon \Reys}{105} \Big) \cos 2\phi 
+\frac{\epsilon \mbox{St}}{15} + \frac{11 \epsilon \Reys}{35} \Big] \,.
\end{align}  
\end{subequations}
\end{widetext}
This equation 
is equivalent to the near-spherical limit 
of Eq.~(9) in Ref.~\cite{einarsson2015a}. 
The derivation outlined in the present article differs 
from the calculations in Refs.~\cite{einarsson2015a,einarsson2015b}
in that the method summarised here relies on a basis expansion in spherical harmonics, and a joint expansion in the particle eccentricity $\epsilon$ and the shear Reynolds number $\Reys$. 
The calculations in Refs.~\cite{einarsson2015a,einarsson2015b}
by contrast make use of a multipole expansion, valid for arbitrary aspect ratios. The fact that the calculations agree 
lends support to the result, to Eq.~(\ref{eq:eom_small_eps}) 
as well as to 
Eq.~(9) in Ref.~\cite{einarsson2015a}. 
We refer the reader to Refs.~\cite{einarsson2015a,einarsson2015b} for a further discussion of the implications of these  results.  

{\em Purely rotational flow.} In the case of a purely rotational flow  the matrix $\tens{A}^\infty$ reads
$$ 
\tens{A}^\infty = \tens{O}^\infty =  \left( \begin{array}{ccc} 0 & -1 & 0 \\ 
1 & 0 & 0\\
0& 0& 0\\
 \end{array} \right) \:,\quad \mbox{so that }  \quad \tens{S}^\infty = \tens{0}\:.
 $$
In this case Jeffery's slip angular velocity {vanishes} and fluid-inertia effects vanish {for small $\Reys$.}
Only particle inertia affects 
the angular velocity of the spheroid, and we find:
\begin{subequations}
\label{eq:rotation_small_eps_1}
\begin{align}    
\dot{\phi} = & 1 \label{eq:phi_rotational}\\
\dot{\theta} = & \frac{\epsilon \mbox{St}}{30} \sin{2\theta} \,.
\label{eq:rotation_small_eps_1b}
\end{align}  
\end{subequations}
In a rotational flow the evolutions of the Euler-angles are thus decoupled. Eq.~(\ref{eq:phi_rotational}) indicates that the {spheroid 
rotates with the same angular velocity as the fluid.}
Eq. (\ref{eq:rotation_small_eps_1b}) admits two equilibrium orientations ({\it modulo} $\pi$)  for the angle $\theta$. The first is  $\theta=0$  (alignment with vorticity), and 
the second is $\theta=\pi/2$ (the particle rotates in the flow plane). Alignment with vorticity is unstable 
for prolate particles ($\epsilon>0$). For oblate nearly spherical particles we find that the stabilities are reversed.

{\em Purely elongational flow.}
In the case of a purely elongational flow,  
the matrix  $\tens{A}^\infty$ reads
\begin{equation}
\tens{A}^\infty = \tens{S}^\infty =  \left( \begin{array}{ccc} 1 & 0 & 0 \\ 
0 & -1 & 0\\
0& 0& 0\\
 \end{array} \right) \:,\quad \mbox{so that }  \quad \tens{O}^\infty = \tens{0}\:.
\nn
\end{equation} 
 In this case  particle-inertia effects vanish so that only  fluid-inertia effects remain:
\begin{subequations}
\label{eq:rotation_small_eps_2}
\begin{align}    
\dot{\phi} = & -{\epsilon} \sin{ 2 \phi} \label{eq:elong_1}\\
\dot{\theta} = &  \frac{{\epsilon}}{2}\sin{2\theta} \cos{2\phi} +  \frac{11 \epsilon \Reys} {70} \sin{2\theta} \label{eq:elong_2}\,.
\end{align}  
\end{subequations}
{As in the rotational flow the} temporal evolution of the azimuthal angle $\phi$  decouples from that of the polar angle $\theta$. Two equilibrium positions are found for the azimuthal angle: $\phi=0$ which is  stable for a prolate particle ($\epsilon > 0$) and unstable for an oblate one ($\epsilon<0$),  and $\phi=\pi/2$ for which the stability is reversed. The equilibrium positions found for $\theta$ are similar to those obtained for $\phi$, that is $\theta=0$ and $\theta=\pi/2$. 
From the stability analysis of the $\phi$-dynamics 
it {follows that} $\epsilon \cos(2\phi)/2$ remains positive for both prolate and oblate particles, so that  only the polar angle $\theta=\pi/2$ is found to be stable. 
As a result,  prolate and oblate particles orient  their axes of symmetry in the flow plane and finally reach fixed orientation repespectively  along $\pm \vec{e}_1$ (prolate) or along $\pm \vec{e}_2$ (oblate). As $\mbox{Re}_s$ is a small parameter, fluid-inertia effects cannot modify these equilibrium  positions. They simply speed up alignment of prolate particles and slow down alignment of oblate particles. 

\section{Conclusions}
In this paper, an equation of motion  has been derived for the orientational dynamics of a neutrally buoyant  and nearly spherical particle,  immersed in a general steady linear flow. {It would be of interest to extend the results to time-dependent linear flows}
of the form $\vec{u}^\infty = \tens{A}^\infty(t) \cdot \vec{x}$. In this case the second part of Eq. (\ref{eq:identities}) does not 
hold any more and {consequently} the torque due to unperturbed flow does not vanish. 
Correction terms scaling as $O(\mbox{Re}_s)$ and $O(\mbox{St})$ arise.  
These additional terms are expected to render the orientational dynamics of the spheroid more complex, {but} 
it remains to be seen in detail how these terms affect the orientational dynamics of a neutrally buoyant particle in an unsteady flow.

\appendix
\section*{Appendix: Stokes flow around a spheroid in a general linear ambient flow}
{In this appendix we describe how to} determine the Stokes flow around a spheroid immersed in a general linear ambient flow. To this end, equations (\ref{eq:Stokes1}) and (\ref{eq:Stokes2}) must be solved. This task is difficult because of the non-spherical shape of the particle. This is the reason why we have used a perturbation method,  briefly described in the following.

The surface of the particle is parametrised by
$$
r(\vartheta) = 1 + \epsilon h(\vartheta) + \epsilon^2 g(\vartheta)
$$
(see Eq.~(\ref{eq:shape}) for the actual definition of the spheroid considered in this study).
We introduce the slip angular velocity tensor
$$
  \boldsymbol{\Omega}_s = \boldsymbol{\Omega} - \tens{O}^\infty\,,\quad \mbox{where} \quad \boldsymbol{\Omega} \cdot \vec{r} = \boldsymbol{\omega}\times \vec{r}\:.
 $$
Performing an expansion to order $O(\epsilon^3)$ of the fluid velocity in the boundary equation {(\ref{eq:Stokes2})} yields
\begin{equation}
\begin{split}
&\vec{u}' + \Big(\epsilon h(\vartheta) + \epsilon^2 g(\vartheta) \Big) \frac{\partial \vec{u}'}{\partial r}\Big|_{r=1} \nn\\
&\quad+ \frac{1}{2} \Big(\epsilon h(\vartheta) + \epsilon^2 g(\vartheta)\Big)^2 \frac{\partial^2 \vec{u}'}{\partial r^2}\Big|_{r=1}  \\
&\quad=  \Big( 1 + \epsilon h(\vartheta) + \epsilon^2 g(\vartheta)\Big) \Big(\boldsymbol{\Omega}_s -\tens{S}^\infty\Big) \cdot \vec{e}_r\:,
\end{split}
\end{equation}
where $\vec{e}_r \equiv \vec{r}/|\vec{r}|$  is the unit radial vector.
We seek a solution of Eqs.~(\ref{eq:Stokes1}) and (\ref{eq:Stokes2}) in the form 
$$
\vec{u}' = \vec{u}'_0 + \epsilon \vec{u}'_1 + \epsilon^2 \vec{u}'_2 + O(\epsilon^3)\:.
$$
Identifying terms of the same order in $\epsilon$ results in 
a set of three sub-problems to solve, associated with the boundary conditions
\begin{eqnarray}
\vec{u}'_0\Big|_{r=1} &=&  \Big(\boldsymbol{\Omega}_s - \tens{S}^\infty \Big) \cdot \vec{e}_r \label{eq:sub1}\\
\vec{u}'_1\Big|_{r=1} 
&= &\Big(\boldsymbol{\Omega}_s - \tens{S}^\infty \Big) \cdot h(\vartheta) \vec{e}_r - h(\vartheta) 
\frac{\partial \vec{u}'_0}{\partial r}\Big|_{r=1}\label{eq:sub2}\\
\vec{u}'_2\Big|_{r=1}  
&= &\Big(\boldsymbol{\Omega}_s - \tens{S}^\infty \Big) \cdot g(\vartheta) \vec{e}_r - g(\vartheta)  \frac{\partial \vec{u}'_0}{\partial r}\Big|_{r=1}\nn \\
& & - h(\vartheta) 
\frac{\partial \vec{u}'_1}{\partial r}\Big|_{r=1}   - \frac{h(\vartheta)^2}{2}\frac{\partial^2 \vec{u}'_0}{\partial r^2}\Big|_{r=1} \label{eq:sub3}\:.
\end{eqnarray}

{\em Lamb solutions.} To determine the solutions of Eqs (\ref{eq:sub1}) to (\ref{eq:sub3}), we use  general Lamb solutions which exploit the fact that any flow solution of the Stokes equations can be written in terms of linear combinations of spherical harmonics.  
According to Lamb \cite{lamb1945} (see also Happel \& Brenner \cite{happel1983}) the general solution for a velocity field that vanishes at infinity can  be cast in the form  
\begin{equation}
\begin{split}
\vec{u}' = & \sum_{n=1}^{\infty} \Big[ \vec{\nabla}
\times \: (r\:\vec{e}_r \:\chi_{-(n+1)}) + \vec{\nabla}
\Phi_{-(n+1)} \\
 &  - \frac{(n - 2)}{ 2n(2n-1)} \:r^2
\vec{\nabla}\:
\mathcal{P}_{-(n+1)}\nn\\
&  + \frac{(n + 1)}{
n(2n-1)} r\:\vec{e}_r\:\mathcal{P}_{-(n+1)}\Big]\:,
\end{split}
\end{equation}
Here $\mathcal{P}_{-(n+1)}$,  $\chi_{-(n+1)}$ and $\Phi_{-(n+1)}$ are spherical harmonics \cite{Byerly1959}. These three spherical harmonics involve  coefficients that remain to be determined using the boundary conditions satisfied by the fluid velocity on the surface of the {particle}. 
Happel and Brenner \cite{happel1983} proposed an efficient method to determine these coefficients that consists in writing the three spherical harmonic functions in the form
$$
\mathcal{P}_{-(n+1)} = \frac{(2n-1)}{(n+1)}
\Big(\frac{1}{r}\Big)^{n+1} \left[(n+2) X_n + Y_n\right]\:,
$$
$$
\Phi_{-(n+1)}= \frac{1}{2(n+1)} \Big(\frac{1}{r}\Big)^{n+1} \left[
n X_n + Y_n \right]\:,
$$
and
$$
\chi_{-(n+1)}= \frac{1}{n(n+1)} \Big(\frac{1}{r}\Big)^{n+1} Z_n\:.
$$
Here $X_n$, $Y_n$ and $Z_n$ are surface harmonics \cite{Abr72}. To give an example, the surface harmonics $X_n$ read
\begin{align}
\label{eq_Xn}
X_n =  &A_n P_n(\cos(\vartheta))\\
 + &\sum_{m=1}^{n} \bigg[B_{nm} \cos(m\varphi) + C_{nm}  \sin(m\varphi) \bigg]  P_n^m(\cos(\vartheta))\,.\nn
\end{align} 
Here $A_{n}$, $B_{nm}$, and $C_{nm}$ are constants to be determined, and $P_n$ and $P_n^m$ are the Legendre polynomials and their  associated functions \cite{Abr72}. 
The surface harmonics $Y_n$ and $Z_n$ have expansions similar to $X_n$ but with different constants. 
The constants are determined by using the relations
\begin{equation}
\vec{e}_r \cdot \vec{V}(\vartheta,\,\varphi) = \sum_{n=1}^{\infty}
X_n(\vartheta,\,\varphi) \label{Xn}\:,
\end{equation}
\begin{equation}
- r \, \vec{\nabla} \cdot \vec{V}(\vartheta,\,\varphi) =
\sum_{n=1}^{\infty} Y_n(\vartheta,\,\varphi) \label{Yn}\:,
\end{equation}
and
\begin{equation}
 r\,
\vec{e}_r \cdot \vec{\nabla} \times \: \vec{V}(\vartheta,\,\varphi)\: =
\sum_{n=1}^{\infty} Z_n(\vartheta,\,\varphi)\:, \label{Zn}
\end{equation}
where by definition
$$
\vec{u}'\Big|_{r=1} =\vec{V}(\vartheta,\,\varphi)\,.
$$
Now the coefficients in $X_n$, $Y_n$ and $Z_n$ are determined as follows.
Let $N$ be the finite number of harmonics in the solution. The value of  $N$ does 
{not exceed  $p+2$ where $p$ is the maximal power of $\cos(\vartheta) $ or $\sin(\vartheta)$ in the boundary condition.} 
The idea here is to  fix first the value of $m$ of the associated Legendre polynomials by exploiting the orthogonality of the trigonometric functions $\cos(m \varphi)$ and $\sin(m \varphi)$ and then to exploit the orthogonality of the Legendre polynomials:
$$
\int_{-1}^{1} P_n(x) P_m(x) \mbox{d}x = \frac{2}{2n+1} \delta_{n\:m}
$$
and their associated functions (for a given $m$): 
$$
\int_{-1}^{1} P_n^m(x)P_{n'}^{m} \mbox{d}x = \frac{2}{2n+1} \frac{(n+m)!}{(n-m)!} \delta_{n\:n'}\:.
$$

As an example, let us show how we determine the coefficients $A_n$, $B_{nm}$ and $C_{nm}$ involved in the surface harmonic $X_n$ in Eq.~(\ref{eq_Xn}) from Eq.~(\ref{Xn}).
{Consider first the case $m=0$. Then:}
$$
T_1 = \frac{1}{2\pi} \int_0^{2\pi}  \Big(\vec{e}_r \cdot \vec{V}(\vartheta,\,\varphi) \Big) \mbox{d}\varphi\,,
$$
and we find for $n$ from $0$ to $N$
$$
A_n = \frac{2 n+1}{2}  \int_{0}^{\pi} \sin(\vartheta) \:P_n(\cos(\vartheta)) \:T_1 \mbox{d}\vartheta \,.
$$
Now consider $\mbox{m}>0$. Let
\begin{equation}
T_{1m} =\frac{1}{\pi} \int_0^{2\pi}  \Big(\vec{e}_r \cdot \vec{V}(\vartheta,\,\varphi) \Big) \cos(m\varphi)  \mbox{d}\varphi\,,\nn
\end{equation}
and 
\begin{equation}
T_{2m} = \frac{1}{\pi} \int_0^{2\pi}  \Big(\vec{e}_r \cdot \vec{V}(\vartheta,\,\varphi) \Big) \sin(m \varphi)  \mbox{d}\varphi\,.\nn
\end{equation}
Then we find for $n$ from $m$ to $N$:
\begin{equation}
B_{nm} =  \frac{2 n+1}{2} \frac{(n-m)!}{(n+m)!}  \int_0^{\pi}\sin(\vartheta) P_n^m(\cos(\vartheta)) \:T_{1m} \mbox{d}\vartheta\,,\nn
\end{equation}
and 
\begin{equation}
C_{nm} =  \frac{2 n+1}{2} \frac{(n-m)!}{(n+m)!}  \int_0^{\pi}\sin(\vartheta) P_n^m(\cos(\vartheta)) \:T_{2m} \mbox{d}\vartheta\,.
\nn
\end{equation}
We implemented this method to determine the solutions of the three sub-problems associated to the boundary conditions  (\ref{eq:sub1})-(\ref{eq:sub3}) in Maple$^{\textregistered}$.


\begin{thebibliography}{14}
\expandafter\ifx\csname natexlab\endcsname\relax\def\natexlab#1{#1}\fi
\expandafter\ifx\csname bibnamefont\endcsname\relax
  \def\bibnamefont#1{#1}\fi
\expandafter\ifx\csname bibfnamefont\endcsname\relax
  \def\bibfnamefont#1{#1}\fi
\expandafter\ifx\csname citenamefont\endcsname\relax
  \def\citenamefont#1{#1}\fi
\expandafter\ifx\csname url\endcsname\relax
  \def\url#1{\texttt{#1}}\fi
\expandafter\ifx\csname urlprefix\endcsname\relax\def\urlprefix{URL }\fi
\providecommand{\bibinfo}[2]{#2}
\providecommand{\eprint}[2][]{\url{#2}}

\bibitem[{\citenamefont{Einarsson
  et~al.}(2015{\natexlab{a}})\citenamefont{Einarsson, Candelier, Lundell,
  Angilella, and Mehlig}}]{einarsson2015a}
\bibinfo{author}{\bibfnamefont{J.}~\bibnamefont{Einarsson}},
  \bibinfo{author}{\bibfnamefont{F.}~\bibnamefont{Candelier}},
  \bibinfo{author}{\bibfnamefont{F.}~\bibnamefont{Lundell}},
  \bibinfo{author}{\bibfnamefont{J.~R.} \bibnamefont{Angilella}},
  \bibnamefont{and} \bibinfo{author}{\bibfnamefont{B.}~\bibnamefont{Mehlig}},
  \bibinfo{journal}{Phys. Rev. E} \textbf{\bibinfo{volume}{91}},
  \bibinfo{pages}{041002} (\bibinfo{year}{2015}{\natexlab{a}}).

\bibitem[{\citenamefont{Einarsson
  et~al.}(2015{\natexlab{b}})\citenamefont{Einarsson, Candelier, Lundell,
  Angilella, and Mehlig}}]{einarsson2015b}
\bibinfo{author}{\bibfnamefont{J.}~\bibnamefont{Einarsson}},
  \bibinfo{author}{\bibfnamefont{F.}~\bibnamefont{Candelier}},
  \bibinfo{author}{\bibfnamefont{F.}~\bibnamefont{Lundell}},
  \bibinfo{author}{\bibfnamefont{J.~R.} \bibnamefont{Angilella}},
  \bibnamefont{and} \bibinfo{author}{\bibfnamefont{B.}~\bibnamefont{Mehlig}},
  \bibinfo{journal}{submitted to Phys. Fluids; arxiv:1504.02309}
  (\bibinfo{year}{2015}{\natexlab{b}}).

\bibitem[{\citenamefont{Lorentz}(1896)}]{lorentz}
\bibinfo{author}{\bibfnamefont{H.}~\bibnamefont{Lorentz}},
  \bibinfo{journal}{Versl. Kon. Akad. Wetensch. Amsterdam}
  \textbf{\bibinfo{volume}{4}}, \bibinfo{pages}{176} (\bibinfo{year}{1896}).

\bibitem[{\citenamefont{Happel and Brenner}(1983)}]{happel1983}
\bibinfo{author}{\bibfnamefont{J.}~\bibnamefont{Happel}} \bibnamefont{and}
  \bibinfo{author}{\bibfnamefont{H.}~\bibnamefont{Brenner}},
  \emph{\bibinfo{title}{Low {R}eynolds number hydrodynamics}}
  (\bibinfo{publisher}{Kluwer Acad. Publisher}, \bibinfo{year}{1983}).

\bibitem[{\citenamefont{Kim and Karrila}(1991)}]{kim1991}
\bibinfo{author}{\bibfnamefont{S.}~\bibnamefont{Kim}} \bibnamefont{and}
  \bibinfo{author}{\bibfnamefont{S.~J.} \bibnamefont{Karrila}},
  \emph{\bibinfo{title}{Microhydrodynamics: principles and selected
  applications}} (\bibinfo{publisher}{Butterworth-Heinemann},
  \bibinfo{address}{Boston}, \bibinfo{year}{1991}).

\bibitem[{\citenamefont{Subramanian and Koch}(2005)}]{subramanian2005}
\bibinfo{author}{\bibfnamefont{G.}~\bibnamefont{Subramanian}} \bibnamefont{and}
  \bibinfo{author}{\bibfnamefont{D.~L.} \bibnamefont{Koch}},
  \bibinfo{journal}{J.~Fluid Mech.} \textbf{\bibinfo{volume}{535}},
  \bibinfo{pages}{383} (\bibinfo{year}{2005}).

\bibitem[{\citenamefont{Subramanian and Koch}(2006)}]{subramanian2006}
\bibinfo{author}{\bibfnamefont{G.}~\bibnamefont{Subramanian}} \bibnamefont{and}
  \bibinfo{author}{\bibfnamefont{D.~L.} \bibnamefont{Koch}},
  \bibinfo{journal}{J.~Fluid Mech.} \textbf{\bibinfo{volume}{557}},
  \bibinfo{pages}{257} (\bibinfo{year}{2006}).

\bibitem[{\citenamefont{Brenner}(1964)}]{Brenner1964}
\bibinfo{author}{\bibfnamefont{H.}~\bibnamefont{Brenner}},
  \bibinfo{journal}{{Chemical Engineering Science}}
  \textbf{\bibinfo{volume}{{19}}}, \bibinfo{pages}{519} (\bibinfo{year}{1964}).

\bibitem[{\citenamefont{Hinch}(1991)}]{Hinch1991}
\bibinfo{author}{\bibfnamefont{E.~J.} \bibnamefont{Hinch}},
  \emph{\bibinfo{title}{Perturbation methods}} (\bibinfo{publisher}{Cambridge
  university press.}, \bibinfo{year}{1991}).

\bibitem[{\citenamefont{Lamb}(1945)}]{lamb1945}
\bibinfo{author}{\bibfnamefont{H.}~\bibnamefont{Lamb}},
  \emph{\bibinfo{title}{Hydrodynamics, 6th edition}}
  (\bibinfo{publisher}{Dover}, \bibinfo{address}{New York},
  \bibinfo{year}{1945}).

\bibitem[{\citenamefont{Jeffery}(1922)}]{jeffery1922}
\bibinfo{author}{\bibfnamefont{G.~B.} \bibnamefont{Jeffery}},
  \bibinfo{journal}{Proceedings of the Royal Society of London. Series A}
  \textbf{\bibinfo{volume}{102}}, \bibinfo{pages}{161} (\bibinfo{year}{1922}).

\bibitem[{\citenamefont{Einarsson et~al.}(2014)\citenamefont{Einarsson,
  Angilella, and Mehlig}}]{Ein14}
\bibinfo{author}{\bibfnamefont{J.}~\bibnamefont{Einarsson}},
  \bibinfo{author}{\bibfnamefont{J.~R.} \bibnamefont{Angilella}},
  \bibnamefont{and} \bibinfo{author}{\bibfnamefont{B.}~\bibnamefont{Mehlig}},
  \bibinfo{journal}{Physica D} \textbf{\bibinfo{volume}{278}},
  \bibinfo{pages}{79} (\bibinfo{year}{2014}).

\bibitem[{\citenamefont{Byerly}(1959)}]{Byerly1959}
\bibinfo{author}{\bibfnamefont{W.~E.} \bibnamefont{Byerly}},
  \emph{\bibinfo{title}{An Elementary Treatise on Fourier Series and Spherical,
  Cylindrical, and Ellipsoidal Harmonics, with Applications to Problems in
  Mathematical Physics}} (\bibinfo{publisher}{Dover. New York},
  \bibinfo{year}{1959}).

\bibitem[{\citenamefont{Abramowitz and Stegun}(1972)}]{Abr72}
\bibinfo{author}{\bibfnamefont{M.}~\bibnamefont{Abramowitz}} \bibnamefont{and}
  \bibinfo{author}{\bibfnamefont{A.}~\bibnamefont{Stegun}},
  \emph{\bibinfo{title}{Handbook of Mathematical Functions, 9th edition}}
  (\bibinfo{publisher}{Dover}, \bibinfo{address}{New York, USA},
  \bibinfo{year}{1972}), \bibinfo{note}{1036p}.

\end{thebibliography}
\end{document}